\documentclass[10pt,twocolumn]{article}
\usepackage[utf8]{inputenc}
\usepackage[T1]{fontenc}
\usepackage{amsmath,amssymb}
\usepackage{graphicx}
\usepackage{hyperref}
\usepackage{booktabs}
\usepackage{xcolor}
\usepackage[margin=2cm]{geometry}
\usepackage{enumitem}
\usepackage{caption}
\usepackage{url}

\hypersetup{colorlinks=true,linkcolor=blue,citecolor=blue,urlcolor=blue}

\title{\textbf{BrainRing: An Interactive Web-Based Tool for Brain Connectivity Chord Diagram Visualization}}

\author{
	Xiao Fan$^{1}$, Yi Zhang$^{1,*}$ \\[6pt]
	$^{1}$Xidian University, Xi'an, China \\[2pt]
	\texttt{xdfanxiao@163.com}
}

\date{}

\begin{document}
\maketitle

\begin{abstract}
	Visualizing brain functional connectivity (FC) patterns is essential for understanding neural organization, yet existing tools such as Circos and BrainNet Viewer require complex configuration files or proprietary software environments. We present \textbf{BrainRing}, a free, open-source, browser-based interactive tool for generating publication-quality chord diagrams of brain connectivity data. BrainRing requires no installation, backend server, or programming knowledge. Users simply open a single HTML file in any modern browser. The tool supports 8 widely-used brain atlases (Brainnetome 246, AAL-90/116, Schaefer 100/200/400, Power 264, and Dosenbach 160), provides real-time parameter adjustment through an intuitive graphical interface, and offers comprehensive edge management including click-to-connect, per-edge color customization, and Circos link file import. BrainRing supports both Chinese and English interfaces and enables researchers to produce publication-ready SVG and PNG figures with full control over visual styling, all within seconds rather than the minutes-to-hours workflow typical of script-based approaches. BrainRing is freely available at \url{https://github.com/XiuFan719/brain-connectivity-viz} with a live demo at \url{https://XiuFan719.github.io/brain-connectivity-viz/}.
\end{abstract}

\section{Introduction}

Functional connectivity (FC) analysis is a cornerstone of modern neuroimaging research, providing insight into how spatially distributed brain regions coordinate their activity \cite{biswal1995functional,fox2007spontaneous}. Among various visualization approaches, chord diagrams (also referred to as Circos plots in the genomics literature \cite{krzywinski2009circos}) have gained popularity for depicting FC patterns. In this representation, brain regions are arranged as arcs along a circular ring, with pairwise connections rendered as curved lines traversing the interior. This layout intuitively captures both the hierarchical organization of brain regions (lobe, gyrus, subregion) and the underlying connectivity structure in a single, compact figure.

Despite their visual appeal, producing publication-quality chord diagrams of brain connectivity remains a cumbersome process. The most commonly used approaches each present notable limitations:

\begin{itemize}[leftmargin=*,nosep]
	\item \textbf{Circos} \cite{krzywinski2009circos}: Originally developed for comparative genomics, Circos requires users to prepare multiple configuration files written in a domain-specific syntax, execute Perl-based rendering scripts, and manually iterate between editing and output inspection. Applying Circos to neuroimaging data further demands the creation of custom karyotype definitions tailored to brain atlases.
	\item \textbf{BrainNet Viewer} \cite{xia2013brainnet}: This tool offers 3D brain surface visualization with connectivity overlays but depends on a licensed MATLAB environment, does not support chord or Circos-style diagram generation, and provides limited interactive control over visual parameters.
	\item \textbf{Custom scripts}: Many researchers resort to writing ad hoc visualization scripts in Python (using matplotlib or MNE-Python) or R. While flexible, this approach requires substantial programming expertise and yields static outputs that must be regenerated from scratch whenever parameters are modified.
\end{itemize}

\begin{table}[h]
	\centering
	\caption{Feature comparison of connectivity visualization tools.}
	\label{tab:comparison}
	\small
	\begin{tabular}{lccc}
		\toprule
		\textbf{Feature} & \textbf{BrainRing} & \textbf{Circos} & \textbf{BNV} \\
		\midrule
		No installation & \checkmark & & \\
		Real-time preview & \checkmark & & \\
		Chord diagram & \checkmark & \checkmark & \\
		3D brain view & & & \checkmark \\
		Built-in atlases & 8 & 0 & 3 \\
		Click-to-edit & \checkmark & & \\
		Per-edge coloring & \checkmark & \checkmark & \\
		Browser-based & \checkmark & & \\
		Bilingual (CN/EN) & \checkmark & & \\
		SVG + PNG export & \checkmark & \checkmark & \\
		\bottomrule
	\end{tabular}
\end{table}

A detailed feature comparison is provided in Table~\ref{tab:comparison}. To address these limitations, we introduce BrainRing, an open-source tool that reduces chord-diagram visualization of brain connectivity to simply opening a webpage. The design of BrainRing is guided by four core principles: (1) \textbf{zero installation}, with the entire application contained in a single HTML file requiring no external dependencies or backend infrastructure; (2) \textbf{real-time interaction}, where every visual parameter is adjustable through sliders and toggles with instant feedback on the canvas; (3) \textbf{atlas-aware design}, with built-in support for 8 major anatomical and functional brain parcellation schemes; and (4) \textbf{publication-ready output}, enabling direct export of figures in both SVG vector format and high-resolution PNG.

\section{Interface Overview}

\begin{figure*}[t]
	\centering
	\includegraphics[width=0.85\textwidth]{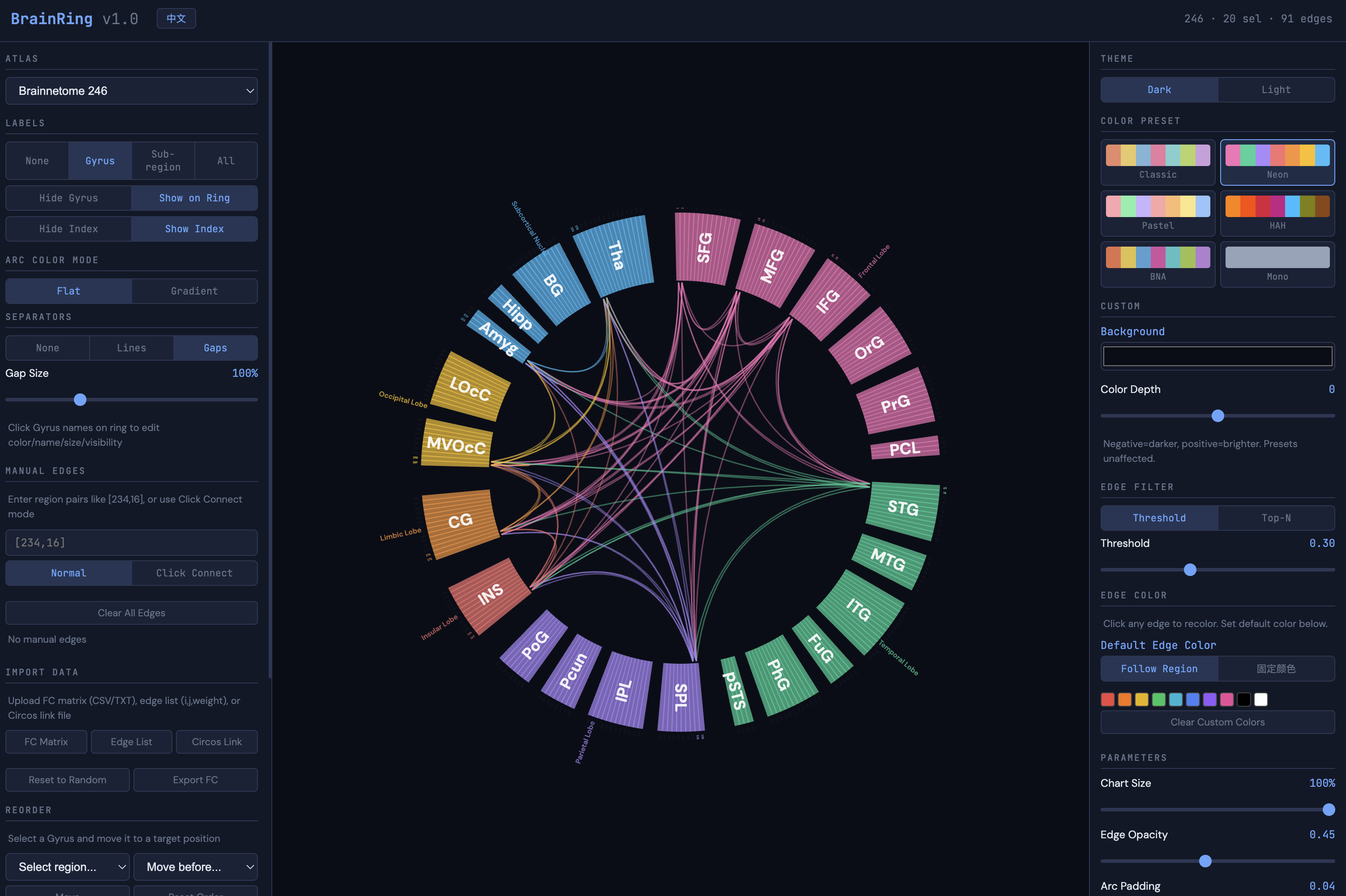}
	\caption{\textbf{BrainRing interface overview.} The application features a three-panel layout: (Left) atlas selection, label configuration, arc color mode, separator settings, manual edge input with click-to-connect mode, data import (FC matrix, edge list, Circos link file), region reorder, and hierarchical region browser; (Center) interactive chord diagram canvas with real-time rendering of the Brainnetome 246 atlas using the Neon color scheme with gap separators; (Right) theme toggle, 6 color presets (including HAH for Circos-compatible per-region coloring), custom background and color depth controls, edge filtering (threshold or Top-N), edge color management, and parameter sliders for ring width, arc padding, line width, font size, rotation, and chart size. A bilingual toggle (CN/EN) is located in the header bar.}
	\label{fig:interface}
\end{figure*}

Figure~\ref{fig:interface} shows the complete BrainRing interface in English mode. The three-panel layout separates data configuration (left), visualization (center), and visual styling (right), allowing researchers to iterate rapidly on figure design without context switching. All parameter changes take effect immediately, with no ``run'' button or compilation step required.

\section{System Design}

\subsection{Architecture}

BrainRing is implemented as a self-contained single-file HTML application ($\sim$185 KB) using D3.js v7 for SVG rendering. All atlas data, color schemes, internationalization dictionaries, and application logic are embedded in the file, enabling fully offline use without any server dependency. The application uses a reactive rendering pipeline: any parameter change triggers a complete re-render of the SVG chord diagram, typically completing in under 50ms on modern hardware. The entire interface supports bilingual operation in Chinese and English, switchable at any time via a toggle button in the header bar, making the tool accessible to both Chinese-speaking and international researchers.

\subsection{Supported Atlases}

BrainRing includes 8 pre-built brain atlas templates spanning both anatomical and functional parcellation schemes (Table~\ref{tab:atlases}).

\begin{table}[h]
	\centering
	\caption{Supported brain atlases in BrainRing.}
	\label{tab:atlases}
	\small
	\begin{tabular}{lrcl}
		\toprule
		\textbf{Atlas} & \textbf{ROIs} & \textbf{Type} & \textbf{Hierarchy} \\
		\midrule
		Brainnetome & 246 & Anat. & Lobe$\rightarrow$Gyrus$\rightarrow$Sub \\
		AAL-90 & 90 & Anat. & Lobe$\rightarrow$Region \\
		AAL-116 & 116 & Anat. & Lobe$\rightarrow$Region \\
		Schaefer-100 & 100 & Func. & Yeo7$\rightarrow$Parcel \\
		Schaefer-200 & 200 & Func. & Yeo7$\rightarrow$Parcel \\
		Schaefer-400 & 400 & Func. & Yeo7$\rightarrow$Parcel \\
		Power-264 & 264 & Func. & 14 Networks \\
		Dosenbach-160 & 160 & Func. & 7 Networks \\
		\bottomrule
	\end{tabular}
\end{table}

\subsection{Data Input}

BrainRing provides multiple flexible approaches for specifying connectivity data, accommodating both automated pipelines and manual curation workflows.

\textbf{FC matrix upload.} Users can upload a full $N \times N$ symmetric functional connectivity matrix in CSV, TXT, or TSV format, where $N$ matches the number of regions in the currently selected atlas. The tool automatically detects the delimiter (comma, tab, or space) and parses the matrix. Upon loading, the threshold is automatically adjusted to display a meaningful number of edges, and all regions involved in supra-threshold connections are selected for display.

\textbf{Edge list upload.} For sparse representations, users can upload edge lists with one connection per line in \texttt{i, j, weight} format, where \texttt{i} and \texttt{j} are 1-indexed region identifiers. This format is convenient for outputs from graph-theoretic analysis pipelines such as the Brain Connectivity Toolbox \cite{rubinov2010complex}.

\textbf{Circos link file import.} BrainRing natively supports the standard Circos link file format (\texttt{hs$i$ start end hs$j$ start end color=$r$,$g$,$b$ value=$v$}), enabling researchers to directly reuse existing Circos configurations without any preprocessing. Per-edge colors specified via the \texttt{color=r,g,b} attribute are automatically parsed and applied to individual connections, preserving the visual encoding from prior Circos-based workflows.

\textbf{Manual edge specification.} In addition to file-based input, users can manually specify connections by typing region index pairs (e.g., \texttt{[234,16]}) into an input field, with batch input supported (e.g., \texttt{[1,2],[3,4],[100,200]}). A dedicated \emph{click-to-connect} mode further simplifies manual edge creation: users click a source region arc on the ring, then click a target region, and the connection is instantly created. The mode automatically resets after each pair, ready for the next connection. All manually added edges are displayed in a management panel showing full region labels (e.g., ``\texttt{[33,69] IFG\_L\_6\_3 $\leftrightarrow$ STG\_L\_6\_1}'') with individual delete buttons, enabling fine-grained curation of the displayed connectivity pattern.

\section{Features}

\subsection{Visual Customization}
\begin{figure*}[t]
	\centering
	\includegraphics[width=\textwidth]{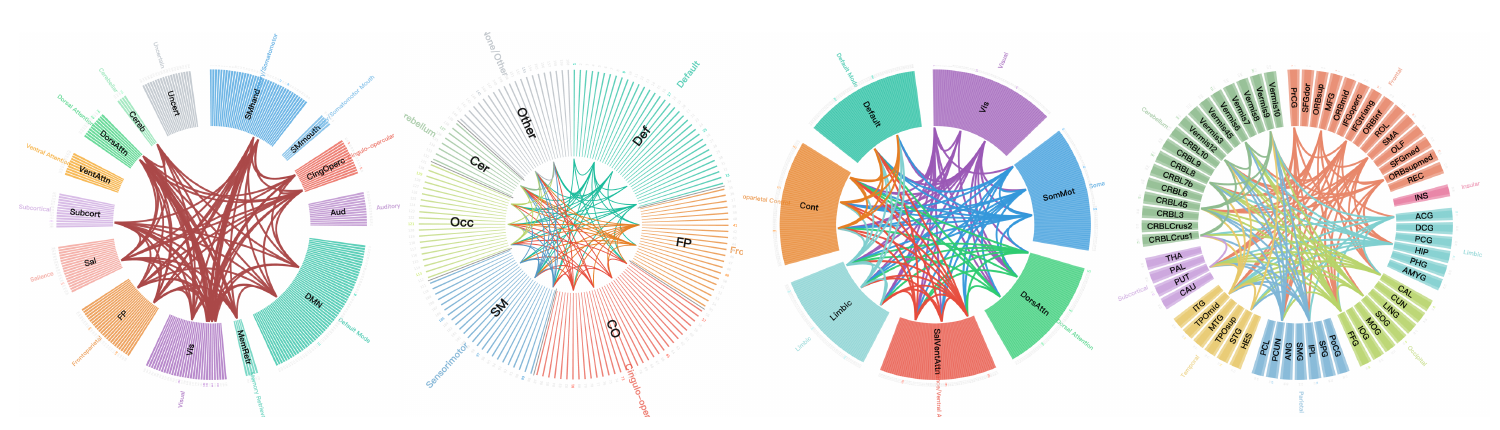}
	\caption{\textbf{Representative outputs generated with BrainRing.} Four examples illustrating different visual configurations.}
	\label{fig:showcase}
\end{figure*}

\textbf{Color control.} Six built-in color schemes are provided, including HAH which reproduces 246 unique per-subregion color assignments from standard Circos neuroimaging templates. A color intensity slider ($-80$ to $+80$) adjusts overall brightness. Arc colors can be rendered in flat (uniform per lobe) or gradient mode (smooth variation within each lobe).

\textbf{Layout parameters.} Ring width (4--100px), arc padding, separator mode (none, lines, or gaps with adjustable size 20\%--300\%), rotation angle (0\textdegree--360\textdegree), and chart size (30--100\%) are continuously adjustable. Arc stroke width and color can be independently controlled for crisp boundary delineation.

\textbf{Labeling.} Four label levels (Lobe, Gyrus, Subregion, Index) can be toggled independently. Gyrus names render directly on the ring with globally or individually adjustable font sizes. Individual Gyrus labels can be clicked to rename, recolor, resize, or hide.

\textbf{Theme.} A dark/light theme toggle affects only the canvas background, allowing white-background export for publications while maintaining a dark panel UI for comfortable editing.

\subsection{Edge Management}

Edges can be filtered by FC threshold or by Top-$N$ selection (strongest, weakest, or absolute). Users can manually add edges by typing region pairs (e.g., \texttt{[234,16]}) or using a dedicated click-to-connect mode where clicking two regions sequentially creates an edge, then automatically resets for the next pair. Clicking any rendered connection opens a color picker with preset swatches and custom input. All manual edges are listed with full region labels and individual delete buttons.

\subsection{Region Reordering}

BrainRing supports Gyrus-level reordering: users select a Gyrus group and move it to any position on the ring. The FC matrix, selection state, manual edges, and custom edge colors are all automatically remapped to maintain data consistency.

\subsection{Internationalization}

The interface supports Chinese and English, switchable via a header button. All 79 static UI elements and 36 dynamic text strings are translated, ensuring accessibility for both Chinese-speaking and international researchers.

\subsection{Export}

Figures can be exported as SVG (vector, infinite resolution) or PNG (rasterized at 4$\times$ resolution). Exported files include the current background color and all visual settings, producing publication-ready output without post-processing.

\section{Showcase}

Figure~\ref{fig:showcase} presents four representative outputs generated with BrainRing, demonstrating different atlas, color scheme, and layout configurations. All examples use a white background for direct inclusion in journal manuscripts.

\section{Availability}

BrainRing is released under the MIT License. The source code and documentation are hosted at \url{https://github.com/XiuFan719/brain-connectivity-viz}, and a live demo is available at \url{https://XiuFan719.github.io/brain-connectivity-viz/}. For questions, suggestions, or collaboration inquiries, please contact \texttt{xdfanxiao@163.com}.
The tool requires only a modern web browser and works on desktop, laptop, and tablet devices.

\section{Conclusion}

BrainRing provides a zero-installation, browser-based solution for interactive brain connectivity chord diagram visualization. By combining real-time parameter adjustment, multi-atlas support, comprehensive edge management, bilingual interface, and publication-quality export in a single HTML file, it substantially lowers the barrier to producing informative connectivity figures.

Beyond neuroimaging, chord diagrams are widely adopted across diverse scientific domains, including genomics and comparative genome analysis \cite{krzywinski2009circos}, microbiome co-occurrence networks, gene regulatory interaction mapping, metabolic pathway visualization, ecological food web analysis, and social network research. The interactive, configuration-free design philosophy of BrainRing can be readily extended to serve these communities. We plan to develop a customizable template system in future releases, allowing users to define their own region hierarchies, color schemes, and label structures for arbitrary circular layout applications beyond brain connectivity.

Researchers with specialized visualization requirements or domain-specific atlas integration needs are welcome to contact us or contribute via the GitHub repository. We believe that lowering the technical barrier to high-quality scientific visualization benefits the broader research community, and we are committed to the continued development and maintenance of BrainRing as an open-source tool.
\bibliographystyle{plain}

\end{document}